\documentclass[twocolumn,showpacs,prb]{revtex4-1}%
\usepackage{amsfonts}
\usepackage{amsmath}
\usepackage{amssymb}
\usepackage{subfigure}
\usepackage{graphicx}
\usepackage{txfonts}%
\providecommand{\U}[1]{\protect\rule{.1in}{.1in}}

\begin{document}
\title{Perfect transmission at oblique incidence by trigonal warping in graphene P-N\ junctions}
\author{Shu-Hui Zhang$^{1,2}$}
\email{shuhuizhang@mail.buct.edu.cn}
\author{Wen Yang$^{2}$}
\email{wenyang@csrc.ac.cn}
\affiliation{$^{1}$College of Science, Beijing University of Chemical Technology, Beijing,
100029, China}
\affiliation{$^{2}$Beijing Computational Science Research Center, Beijing 100193, China}

\begin{abstract}
We develop an analytical mode-matching technique for the tight-binding model
to describe electron transport across graphene P-N junctions. This method
shares the simplicity of the conventional mode-matching technique for the
low-energy continuum model and the accuracy of the tight-binding model over a
wide range of energies. It further reveals an interesting phenomenon on a
sharp P-N junction: the disappearance of the well-known Klein tunneling (i.e.,
perfect transmission) at normal incidence and the appearance of perfect
transmission at oblique incidence due to trigonal warping at energies beyond
the linear Dirac regime. We show that this phenomenon arises from the
conservation of a generalized pseudospin in the tight-binding model. We expect
this effect to be experimentally observable in graphene and other Dirac
fermions systems, such as the surface of three-dimensional topological insulators.

\end{abstract}

\maketitle

\section{Introduction}

Klein tunneling \cite{KleinZP1929}, the unimpeded penetration of relativistic
particles regardless of the height and width of potential barriers, is an exotic
effect compared with the exponential-decaying transmission of nonrelativistic
particles \cite{AllainEPJB2011}. In 2006, the seminal theoretical work of
Katsnelson \textit{et al}. \cite{KatsnelsonNatPhys2006} brought about the
possibility of demonstrating Klein tunneling across electrostatic junctions in
graphene. This proposal has stimulated widespread theoretical
\cite{CheianovPRB2006,PereiraPRB2006,BaiPRB2007,BeenakkerPRB2008,SetareJPCM2010,RoslyakJPCM2010,ZebPRB2008,SoninPRB2009,SchelterPRB2010,YangPRB2011c,RozhkovPhysRep2011,LiuPRB2012b,GiavarasPRB2012,RodriguezJAP2012,PopoviciPRB2012,HeinischPRB2013,LogemannPRB2015,OhPRL2016,ErementchoukJPCM2016,DowningJPCM2017}
and experimental
\cite{HuardPRL2007,GorbachevNL2008,StanderPRL2009,YoungNatPhys2009,RossiPRB2010,SajjadPRB2012,SutarNL2012,RahmanAPL2015,GutierrezNatPhys2016,ChenScience2016,LaitinenPRB2016,BaiPRB2017}
interest and shed light on possible electronic applications
\cite{SajjadAPL2011,JangPNAS2013,Wilmart2DM2014,ChenPRB2015,SousaJAP2017,Tan2017}.

\begin{figure}[t]
\includegraphics[width=\columnwidth,clip]{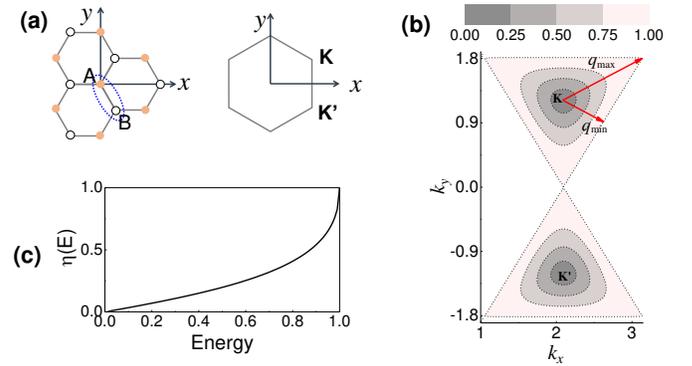}\caption{(a)
Honeycomb lattice of graphene in the real space (left, the dashed ellipse for a
unit cell) and reciprocal space (right). (b) Constant energy contour in the
two inequivalent valleys: $\mathbf{K}$ and $\mathbf{K}^{\prime}$. (c) Degree
of trigonal warping $\eta(E)\equiv[q_{\max}(E)-q_{\min}(E)]/q_{\min}(E)$ as a
function of energy $E$, where $q_{\mathrm{max}}(E)$ [$q_{\mathrm{min}}(E)$] is
the largest (smallest) reduced wave vector on the constant energy contour $E$
[see panel (b)].}%
\label{G_GRAPHENE}%
\end{figure}

Graphene is a 2D layer of carbon atoms on a honeycomb lattice [Fig.
\ref{G_GRAPHENE}(a)]. The conduction band and the valence band touch each
other at six Dirac points, but only two of them are inequivalent, as denoted
by $\mathbf{K}$ and $\mathbf{K}^{\prime}$ in the right panel of Fig.
\ref{G_GRAPHENE}(a). According to group theory \cite{PikusStrainBook1974},
the D$_{\mathrm{6h}}$ point group symmetry of the honeycomb lattice determines
the global\textit{ }hexagonal symmetry of the graphene energy band over the
first Brillouin zone, while the local D$_{\mathrm{3h}}$ symmetry of each
valley determines the local triangular symmetry of the energy band in this
valley. Close to the Dirac point, say $\mathbf{K}$, the energy dispersion as a
function of the reduced wave vector\textit{ }$\mathbf{q}\equiv\mathbf{k}%
-\mathbf{K}$ is linear and isotropic. Away from the Dirac point, however, the
D$_{\mathrm{3h}}$ local symmetry becomes important and the constant energy
contour approaches a regular triangle with a side length $2\pi/3$ [see Fig.
\ref{G_GRAPHENE}(b)]. This is commonly referred to as \textit{trigonal
warping}, which increases rapidly with energy [Fig. \ref{G_GRAPHENE}(c)]. Note
that throughout this work we\textit{ }use the carbon-carbon bond length
$a=0.142$ nm as the unit of length and the nearest-neighbor hopping energy
$t_{0}=2.7$ eV as the unit of energy \cite{CastroRMP2009}.

Trigonal warping is a well-known feature of the graphene energy band beyond
the linear regime \cite{CastroRMP2009} and its influence on the electron
transport is receiving growing interest, including the observation of broken
chirality by trigonal warping in transport measurements
\cite{WuPRL2007a,DombrowskiPRL2017}, the influence
\cite{XingPRB2010,ReijndersPRB2017} of trigonal warping on the famous Veselago
focusing across graphene P-N junctions
\cite{CheianovScience2007,LeeNatPhys2015,ChenScience2016}, and potential
applications of trigonal warping for producing valley-polarized electrons in
N-P-N junction \cite{GarciaPomarPRL2008}, double barriers \cite{JrJPCM2009},
and other junctions \cite{LogemannPRB2015}. To incorporate trigonal warping,
the simplest method \cite{GarciaPomarPRL2008,JrJPCM2009} is to introduce
nonlinear corrections to the widely used low-energy linear continuum model.
The nearest-neighbor tight-binding model captures the hexagonal lattice
symmetry and hence is commonly used
\cite{CastroRMP2009,XingPRB2010,LogemannPRB2015,ReijndersPRB2017} to study the
trigonal warping effect over a wide energy range. At energies below 1 eV, this model gives very accurate energy bands, but its deviation from the \textit{ab initio} calculation becomes significant at energies approaching the Van Hove singularity at $\sim3$ eV \cite{ReichPRB2002}.
However, transport calculations in the tight-binding model are usually based on
the recursive Green's function method
\cite{DattaBook1995,FerryBook1997,LewenkopfJCE2013,ZhangPRB2017} and hence are
numerical. Moreover, previous studies on Klein tunneling mainly focus on the
low-energy linear regime of graphene, leaving the effect of trigonal warping
largely unexplored. In particular, it is well known that the Klein tunneling
(i.e., perfect transmission) at normal incidence originates from the
conservation of the pseudospin \cite{KatsnelsonNatPhys2006,BeenakkerRMP2008}.
However, this simple physical picture is based on the linear continuum model
and hence is limited to the low-energy linear Dirac regime. It is not clear
whether a similar physical picture exists in the tight-binding model and over
a wide range of energies.

In this paper, we develop an analytical mode-matching technique in the
tight-binding model for electron transport across graphene P-N junctions. The
key is to introduce a titled coordinate system, reduce the 2D junction into a
1D chain, and then perform mode-matching at the P-N interface, in a similar
way to the mode matching in the continuum model
\cite{FerryBook1997,KatsnelsonNatPhys2006}. As a result, this method shares
the simplicity of the mode-matching method in the low-energy continuum model
and the accuracy of the tight-binding model over a wide range of energies. To
focus on trigonal warping, we consider the electron transmission across a
sharp P-N interface along the zigzag direction in the energy range
$E\in[-1,+1]$, where the intervalley scattering is absent
\cite{GarciaPomarPRL2008}; but trigonal warping is significant and can be
described reasonably by the tight-binding model. The results show that beyond
the linear regime, the Klein tunneling at normal incidence becomes imperfect;
i.e., finite backscattering occurs. Interestingly, we find that perfect
transmission is still possible, but the critical incident angle for perfect
transmission deviates from zero, and the deviation increases with increasing
trigonal warping. We introduce the concept of a generalized pseudospin in the
tight-binding model and show that its conservation across the P-N interface is
responsible for the perfect transmission at oblique incidence. This
generalizes the well-known pseudospin picture for perfect transmission,
previously limited to the linear Dirac regime, to a wide energy range
$E\in[-1,+1]$. The continuum model with a second-order nonlinear correction
fails to describe this phenomenon quantitatively. We expect this phenomenon to
be experimentally observable.

The rest of this paper is organized as follows. In Sec. II, we introduce the
tilted coordinates, present the analytical mode-matching method, and introduce
the generalized pseudospin picture for perfect transmission. In Sec. III, we
discuss the perfect transmission at oblique incidence and the failure of the
continuum model. Finally, we give a brief conclusion in Sec. IV.

\section{Mode-matching technique in tilted coordinates}

\begin{figure}[t]
\includegraphics[width=\columnwidth,clip]{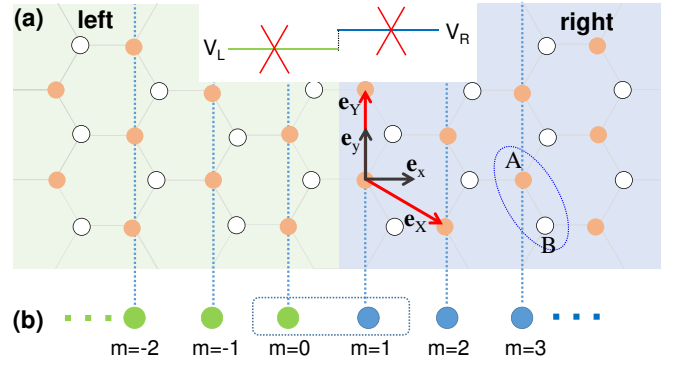}\caption{(a) Graphene
junction with a zigzag interface. The inset sketches the Dirac points of the
left region and the right region. (b) One-dimensional chain along the $X$
axis, with two orbitals $A$ and $B$ in one unit cell.}%
\label{G_PNJ}%
\end{figure}

For specificity, we consider a graphene P-N\ junction with a sharp, zigzag
interface separating the left region and the right region [Fig. \ref{G_PNJ}%
(a)], leaving the more general cases to the end of this section. The
tight-binding Hamiltonian of the junction is the sum of the
(nearest-neighbor)\ tight-binding Hamiltonian for uniform graphene
\cite{CastroRMP2009} and the on-site junction potential, which takes a
constant value $V_{\mathrm{L}}$ ($V_{\mathrm{R}}$) for all the carbon sites in
the left (right) region [see the inset of Fig. \ref{G_PNJ}(a)]. The Fermi
level $E_{F}$, which can be tuned by electric gating, determines the doping
level in the left (right) region as $\varepsilon_{\mathrm{L}}\equiv
E_{F}-V_{\mathrm{L}}$ ($\varepsilon_{\mathrm{R}}\equiv E_{F}-V_{\mathrm{R}}$),
where a positive (negative) doping level means electron or N (hole or P)
doping. Here we consider a P-N\ junction with $\varepsilon_{\mathrm{L}}>0$ and
$\varepsilon_{\mathrm{R}}<0$; i.e., the left region is N\ doped and the right
region is P doped.

In addition to the ortho-normalized basis vectors $(\mathbf{e}_{x}%
,\mathbf{e}_{y})$ of the conventional Cartesian coordinate system, we
introduce a tilted coordinate system characterized by the nonorthogonal basis
vectors $(\mathbf{e}_{X},\mathbf{e}_{Y})$ [see Fig. \ref{G_PNJ}(b)]. The
tilted vectors $\mathbf{e}_{X}$ and $\mathbf{e}_{Y}$ are actually the two
primitive vectors for the Bravais lattice of uniform graphene and are
connected to $(\mathbf{e}_{x},\mathbf{e}_{y})$ by%
\[
\left\{
\begin{array}
[c]{l}%
\mathbf{e}_{X}=\sqrt{3}(\dfrac{\sqrt{3}}{2}\mathbf{e}_{x}-\dfrac{1}%
{2}\mathbf{e}_{y}),\\
\mathbf{e}_{Y}=\sqrt{3}\mathbf{e}_{y},
\end{array}
\right.  \Leftrightarrow\left\{
\begin{array}
[c]{l}%
\mathbf{e}_{x}=\dfrac{2\mathbf{e}_{X}+\mathbf{e}_{Y}}{3},\\
\mathbf{e}_{y}=\dfrac{\mathbf{e}_{Y}}{\sqrt{3}}.
\end{array}
\right.
\]
The Cartesian components $k_{x}\equiv\mathbf{k}\cdot\mathbf{e}_{x}$ and
$k_{y}\equiv\mathbf{k}\cdot\mathbf{e}_{y}$ of a wave vector $\mathbf{k}$ in
the reciprocal space are connected to the tilted components $k_{X}%
\equiv\mathbf{k}\cdot\mathbf{e}_{X}$ and $k_{Y}\equiv\mathbf{k}\cdot
\mathbf{e}_{Y}$ via
\begin{equation}
\left\{
\begin{array}
[c]{l}%
k_{X}=\sqrt{3}(\dfrac{\sqrt{3}}{2}k_{x}-\dfrac{1}{2}k_{y}),\\
k_{Y}=\sqrt{3}k_{y},
\end{array}
\right.  \Leftrightarrow\left\{
\begin{array}
[c]{l}%
k_{x}=\dfrac{2k_{X}+k_{Y}}{3},\\
k_{y}=\dfrac{k_{Y}}{\sqrt{3}}.
\end{array}
\right.  \label{KXKY}%
\end{equation}
For a position vector $\mathbf{R}$ in the real space, we define the tilted
components $R_{X},R_{Y}$ through the expansion%
\[
\mathbf{R}\equiv R_{X}\mathbf{e}_{X}+R_{Y}\mathbf{e}_{Y}\Rightarrow\left\{
\begin{array}
[c]{l}%
R_{X}=\dfrac{2R_{x}}{3},\\
R_{Y}=\dfrac{R_{x}+\sqrt{3}R_{y}}{3}.
\end{array}
\right.
\]
These definitions lead to the convenient expression $\mathbf{k}\cdot
\mathbf{R}=k_{x}R_{x}+k_{y}R_{y}=k_{X}R_{X}+k_{Y}R_{Y}$, although
$\mathbf{k}\neq k_{X}\mathbf{e}_{X}+k_{Y}\mathbf{e}_{Y}$ and $R_{X,Y}%
\neq\mathbf{R}\cdot\mathbf{e}_{X,Y}$ since $(\mathbf{e}_{X},\mathbf{e}_{Y})$
are not ortho-normalized. An arbitrary vector $\mathbf{O}$ can be denoted by
$\mathbf{O}=(O_{x},O_{y})=(O_{X},O_{Y})_{\mathrm{T}}$.

\subsection{Tight-binding model for uniform graphene}

Here we consider uniform graphene with $\varepsilon_{\mathrm{L}}%
=\varepsilon_{\mathrm{R}}=E_{F}$ to illustrate the usage of the tilted
coordinates and establish the relevant notations and some important concepts.

As shown in Fig. \ref{G_GRAPHENE}(a) or \ref{G_PNJ}(a), the honeycomb lattice
of graphene consists of two sublattices (denoted by $A$ and $B$) and each unit
cell contains two carbon atoms, one on each sublattice. The unit cell $(m,n)$
locates at $\mathbf{R}_{m,n}=m\mathbf{e}_{X}+n\mathbf{e}_{Y}=(m,n)_{\mathrm{T}%
}$, and the relative displacements of the two carbon atoms inside this unit
cell are $\boldsymbol{\tau}_{A}=0$ and $\boldsymbol{\tau}_{B}=(1/2,-\sqrt
{3}/2)$. The tight-binding Hamiltonian of uniform graphene is
\[
\hat{H}_{0}=-\sum_{m,n}(|m+1,n,A\rangle+|m,n-1,A\rangle+|m,n,A\rangle)\langle
m,n,B|-h.c.,
\]
where $|m,n,\lambda\rangle$ ($\lambda=A,B$) denotes the carbon atom on the
sublattice $\lambda$. To utilize the translational invariance of graphene with
the primitive vector $\mathbf{e}_{Y}$ along the $Y$ axis, we make a Fourier
transform from the on-site basis $\{|m,n,\lambda\rangle\}$ to the hybrid basis
(i.e., on-site basis along the $X$ axis and Bloch basis along the $Y$ axis)
\begin{equation}
|m,k_{Y},\lambda\rangle=\frac{1}{\sqrt{N_{Y}}}\sum_{n}e^{i{n}k_{Y}%
}|m,n,\lambda\rangle, \label{FT_X}%
\end{equation}
where $N_{Y}$ is the normalization length along the $Y$ axis and $k_{Y}%
\in\lbrack0,2\pi]$. Under this basis $\hat{H}_{0}=\sum_{k_{Y}}\hat{h}(k_{Y})$,
where
\[
\hat{h}(k_{Y})=-\sum_{m}[(1+e^{ik_{Y}})|m,k_{Y},A\rangle+|m+1,k_{Y}%
,A\rangle]\langle m,k_{Y},B|-h.c..
\]
For a given $k_{Y}$, the Hamiltonian $\hat{h}(k_{Y})$ describes a 1D chain
along the $X$ axis [see Fig. \ref{G_PNJ}(b)], with two orbitals $A$ and $B$ in
each unit cell. The 2$\times$2 on-site energy of each unit cell is
\begin{equation}
\mathbf{h}(k_{Y})=-\left[
\begin{array}
[c]{cc}%
0 & 1+e^{ik_{Y}}\\
1+e^{-ik_{Y}} & 0
\end{array}
\right]  \label{HKY}%
\end{equation}
and the 2$\times$2 hopping matrix from unit cell $m+1$ to $m$ is
\begin{equation}
\mathbf{t}\equiv\langle m|\hat{h}(k_{Y})|m+1\rangle=%
\begin{bmatrix}
0 & 0\\
-1 & 0
\end{bmatrix}
. \label{T}%
\end{equation}
So the Fourier transform reduces the 2D graphene into decoupled 1D chains
[Fig. \ref{G_PNJ}(b)] parametrized by different $k_{Y}$'s. \begin{figure}[pth]
\includegraphics[width=\columnwidth,clip]{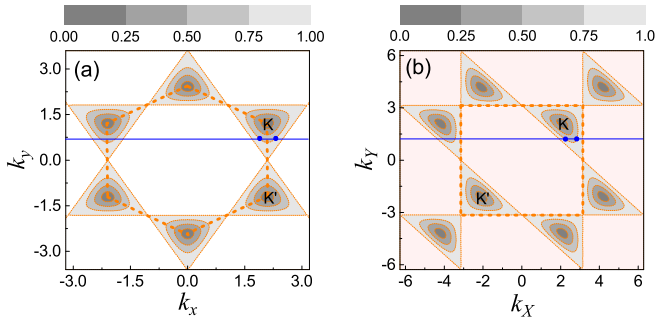}\caption{Fermi contours
$E_{F}=0.25$, $0.5$, $0.75$, and $1.0$ for the conduction band of uniform
graphene in Cartesian (a) and tilted (b) coordinate system. The dashed orange
lines mark the first Brillouin zone. The horizontal blue line corresponds to
$k_{y}=0.70$ in (a) and $k_{Y}=\sqrt{3}k_{y}=1.21$ in (b).}%
\label{G_FBZ}%
\end{figure}

Next, to utilize the translational invariance along the $X$ axis with the
primitive vector $\mathbf{e}_{X}$, we make another Fourier transform along $X$
to get the full Bloch basis:
\[
|\mathbf{k},\lambda\rangle\equiv\frac{1}{\sqrt{N_{X}}}\sum_{m}e^{i{m}k_{X}%
}|m,k_{Y},\lambda\rangle=\frac{1}{\sqrt{N}}\sum_{m,n}e^{i\mathbf{R}_{m,n}%
\cdot\mathbf{k}}|m,n,\lambda\rangle
\]
parameterized by the wave vector $\mathbf{k}=(k_{X},k_{Y})_{\mathrm{T}}$. This
transformation further decouples the lattice degree of freedom along the $X$
axis and leads to
\[
\hat{H}_{0}=\sum_{\mathbf{k}}[|\mathbf{k},A\rangle,|\mathbf{k},B\rangle
]\mathbf{H}_{0}(\mathbf{k})%
\begin{bmatrix}
\langle\mathbf{k},A|\\
\langle\mathbf{k},B|
\end{bmatrix}
,
\]
where
\begin{equation}
\mathbf{H}_{0}(\mathbf{k})\equiv%
\begin{bmatrix}
0 & f(\mathbf{k})\\
f(-\mathbf{k}) & 0
\end{bmatrix}
\label{H0K}%
\end{equation}
is the 2$\times$2 tight-binding Hamiltonian and
\begin{equation}
f(\mathbf{k})=-(1+e^{ik_{Y}%
}+e^{-ik_{X}}).
\label{FK}
\end{equation}

The eigenenergies are $E^{(\pm)}(\mathbf{k})=\pm
|f(\mathbf{k})|$, where
\begin{align}
\left\vert f(\mathbf{k})\right\vert  &  =\sqrt{3+2[\cos k_{X}+\cos k_{Y}%
+\cos(k_{X}+k_{Y})]}\nonumber\\
&  =\sqrt{3+2\cos(\sqrt{3}k_{y})+4\cos\frac{3k_{x}}{2}\cos\frac{\sqrt{3}k_{y}%
}{2}}. \label{ABSFK}%
\end{align}
The corresponding eigenstates are $|\Phi^{(\pm)}(\mathbf{k})\rangle=[1,\pm
f^{\ast}(\mathbf{k})/\left\vert f(\mathbf{k})\right\vert ]^{T}$. Note that we
normalize every spinor to $\sqrt{2}$ throughout this work.

In the Cartesian coordinates, the first Brillouin zone is a regular hexagon,
as shown in Fig. \ref{G_FBZ}(a). In the tilted coordinates, the first
Brillouin zone is a square $k_{X},k_{Y}\in\lbrack-\pi,\pi]$, as shown in Fig.
\ref{G_FBZ}(b). The two inequivalent Dirac points are%
\begin{align*}
\mathbf{K}  &  \equiv\left(  \frac{2\pi}{3},\frac{2\pi}{3\sqrt{3}}\right)
=\left(  \frac{2\pi}{3},\frac{2\pi}{3}\right)  _{\mathrm{T}},\\
\mathbf{K}^{\prime}  &  \equiv\left(  \frac{2\pi}{3},-\frac{2\pi}{3\sqrt{3}%
}\right)  =\left(  \frac{4\pi}{3},-\frac{2\pi}{3}\right)  _{\mathrm{T}%
}\longrightarrow\left(  -\frac{2\pi}{3},-\frac{2\pi}{3}\right)  _{\mathrm{T}},
\end{align*}
where in the last step of the second line we have shifted $\mathbf{K}^{\prime
}$ by a reciprocal vector along the $X$ axis.

The continuum model Hamiltonian is obtained from $\mathbf{H}_{0}(\mathbf{k})$
by considering $\mathbf{k}$ in a given valley, say $\mathbf{K}$, defining the
reduced wave vector $\mathbf{q}\equiv\mathbf{k}-\mathbf{K}$, and expanding
$f(\mathbf{k})$ into Taylor series with respect to $\mathbf{q}$. For example,
expanding $f(\mathbf{k})$ up to the first order of $\mathbf{q}$ gives the
widely used linear continuum model. For the $\mathbf{K}$ valley, we have
$f(\mathbf{k})\approx v_{F}|\mathbf{q}|e^{-i(\pi/6-\theta_{\mathbf{q}})}$ and
hence the linear continuum model \
\begin{equation}
\mathbf{H}_{\mathbf{K}}(\mathbf{q})=v_{F}|\mathbf{q}|%
\begin{bmatrix}
0 & e^{-i(\pi/6-\theta_{\mathbf{q}})}\\
e^{i(\pi/6-\theta_{\mathbf{q}})} & 0
\end{bmatrix}
, \label{HK}%
\end{equation}
where $v_{F}\equiv3/2$ is the Fermi velocity and $\theta_{\mathbf{q}}$ is the
azimuth angle of $\mathbf{q}$ in the Cartesian coordinates. For the
$\mathbf{K}^{\prime}$ valley, $f(\mathbf{k})\approx-v_{F}|\mathbf{q}%
|e^{i(\pi/6-\theta_{\mathbf{q}})}$; thus
\begin{equation}
\mathbf{H}_{\mathbf{K}^{\prime}}(\mathbf{q})=-v_{F}|\mathbf{q}|%
\begin{bmatrix}
0 & e^{i(\pi/6-\theta_{\mathbf{q}})}\\
e^{-i(\pi/6-\theta_{\mathbf{q}})} & 0
\end{bmatrix}
. \label{HKP}%
\end{equation}
The linear continuum model for either valley gives the same isotropic, linear
dispersion $E^{(\pm)}(\mathbf{q})=\pm v_{F}|\mathbf{q}|$ that totally ignores
the trigonal warping. The corresponding eigenstates are $|\Phi_{\mathbf{K}%
}^{(\pm)}(\mathbf{q})\rangle=[1,\pm e^{i(\pi/6-\theta_{\mathbf{q}})}]^{T}$ for
the $\mathbf{K}$ valley and $|\Phi_{\mathbf{K}^{\prime}}^{(\pm)}%
(\mathbf{q})\rangle=[1,\mp e^{-i(\pi/6-\theta_{\mathbf{q}})}]^{T}$ for the
$\mathbf{K}^{\prime}$ valley. By expanding $f(\mathbf{k})$ into high orders of
$\mathbf{q}$, trigonal warping can be included into the continuum model.

Now we discuss a distinguishing feature of the tight-binding model compared
with all the continuum models (including those with high-order corrections for
trigonal warping): the existence of \textquotedblleft
abnormal\textquotedblright\ evanescent states. Let us consider a given $k_{y}$
(or $q_{y}$) and determine the $k_{x}$ (or $q_{x}$) of all the eigenstates on
the Fermi level. For $q_{y}<q_{F}\equiv|E_{F}|/v_{F}$, the linear continuum
model of either valley gives two \textit{traveling }eigenstates characterized
by the reduced wave vector$\ (\pm(q_{F}^{2}-q_{y}^{2})^{1/2},q_{y})$ in the
Cartesian coordinate. For the tight-binding model, $k_{x}$ is determined by
$E_{F}=\mathrm{sgn}(E_{F})|f(\mathbf{k})|$, where $\mathbf{k}\equiv
(k_{x},k_{y})$ and $\mathrm{sgn}(x)\equiv+1$ for $x>0$ and $-1$ for $x<0$. The
solutions can be visualized as the two intersection points between the
horizontal line $k_{y}$ and the Fermi contour $E_{F}=\mathrm{sgn}%
(E_{F})|f(\mathbf{k})|$ in the $(k_{x},k_{y})$ plane; e.g., for $E_{F}=0.75$
and $k_{y}=0.70$, we obtain two traveling eigenstates [blue dots in Fig.
\ref{G_FBZ}(a)]. When $k_{y}$ approaches the Dirac point, the eigenstates from
the tight-binding model approach those from the linear continuum model.
Similarly, in the tilted coordinate system, the eigenstates with a given
$k_{Y}$ on the Fermi level can also be determined as the two intersection
points between the horizontal line $k_{Y}$ and the Fermi contour
$E_{F}=\mathrm{sgn}(E_{F})|f(\mathbf{k})|$ in the $(k_{X},k_{Y})_{\mathrm{T}}$
plane, as shown in Fig. \ref{G_FBZ}(b) for $E_{F}=0.75$ and $k_{Y}=1.21$.

Surprisingly, in addition to these two \textquotedblleft
normal\textquotedblright\ eigenstates, the tight-binding model also has two
extra, \textquotedblleft abnormal\textquotedblright\ eigenstates. To make this
clear, we follow Ref. \onlinecite{ZhangPRB2017} and solve the 1D
Schr\"{o}dinger equation
\begin{equation}
-\mathbf{t}^{\dag}|\Phi(m-1)\rangle+(E-\mathbf{h}(k_{Y})\mathbf{)}%
|\Phi(m)\rangle-\mathbf{t}|\Phi(m+1)\rangle=0 \label{SEQ}%
\end{equation}
for the 1D chain along the $X$ axis under the Bloch condition $|\Phi
(m)\rangle=e^{imk_{X}}|\Phi\rangle$, where $k_{X}$ may be either real or
complex. Equation (\ref{SEQ}) can be written as
\begin{equation}
\mathbf{H}_{0}(\mathbf{k})|\Phi\rangle=E|\Phi\rangle, \label{HEQ}%
\end{equation}
where $\mathbf{k}\equiv(k_{X},k_{Y})_{\mathrm{T}}=(k_{x},k_{y})$ is the wave
vector and $\mathbf{H}_{0}(\mathbf{k})$ is the 2$\times$2 Hamiltonian of the
tight-binding model [see Eq. (\ref{H0K})]. For a \textit{given}, real\textit{
}wave vector $\mathbf{k}$, Eq. (\ref{HEQ}) reproduces the energy dispersions
and eigenstates of uniform graphene. Here we need to find \textit{all} the
eigenstates with a given $k_{Y}$ on the Fermi level. For this purpose, we let
$\lambda\equiv e^{ik_{X}}$ be the propagation phase along the $+X$ axis by one
unit cell, set $E=E_{F}$, and rewrite Eq. (\ref{SEQ}) as%
\begin{equation}
\lbrack-\mathbf{t}^{\dagger}+\lambda(E_{F}-\mathbf{h}(k_{Y})\mathbf{)}%
-\lambda^{2}\mathbf{t}]|\Phi\rangle=0, \label{SEQ2}%
\end{equation}
from which we obtain four solutions for $\lambda$ (and hence $k_{X}$) and
$|\Phi\rangle$. Two solutions correspond to the \textquotedblleft
normal\textquotedblright\ eigenstates and can be explicitly obtained by
rewriting Eq. (\ref{SEQ2}) as an explicit quadratic equation for $\lambda$:%
\[
(1+e^{ik_{Y}})\lambda^{2}+[(3+2\cos k_{Y})-E_{F}^{2}]\lambda+(1+e^{-ik_{Y}%
})=0.
\]
The other two solutions are \textquotedblleft abnormal\textquotedblright%
\ evanescent eigenstates, including a right-going one
\begin{equation}
\lambda_{0}=0,\ \ \ |\Phi_{0}\rangle=%
\begin{bmatrix}
\sqrt{2}\\
0
\end{bmatrix}
\label{EIG0}%
\end{equation}
and a left-going one
\begin{equation}
\lambda_{\infty}=\infty,\ \ |\Phi_{\infty}\rangle=%
\begin{bmatrix}
0\\
\sqrt{2}%
\end{bmatrix}
. \label{EIGINF}%
\end{equation}

Finally we discuss the physical meaning of these \textquotedblleft
abnormal\textquotedblright\ evanescent eigenstates. The wave function of the
right-going one [Eq. (\ref{EIG0})] starting from a unit cell $m_{0}$ is
defined in the half plane on the right of $m_{0}$ only (i.e., $m\geq m_{0}$):
\[
|\Phi_{0}(m|m_{0})\rangle=\lambda_{0}^{m-m_{0}}|\Phi_{0}\rangle=\left\{
\begin{array}
[c]{ll}%
|\Phi_{0}\rangle & (m=m_{0}),\\
0 & (m\geq m_{0}+1).
\end{array}
\right.
\]
The wave function of the left-going one [Eq. (\ref{EIGINF}] starting from a
unit cell $m_{0}$ is defined in the half plane on the left of $m_{0}$ only
(i.e., for $m\leq m_{0}$):%
\[
|\Phi_{\infty}(m|m_{0})\rangle=\lambda_{\infty}^{m-m_{0}}|\Phi_{\infty}%
\rangle=\left\{
\begin{array}
[c]{ll}%
|\Phi_{\infty}\rangle & (m=m_{0}),\\
0 & (m\leq m_{0}-1).
\end{array}
\right.
\]
Using $\mathbf{t}^{\dagger}|\Phi_{0}\rangle=\mathbf{t}|\Phi_{\infty}\rangle
=0$, we can readily verify that $|\Phi_{0}(m|m_{0})\rangle$ ($|\Phi_{\infty
}(m|m_{0})\rangle$) indeed satisfies the Schr\"{o}dinger equation Eq.
(\ref{SEQ}) with $m\geq m_{0}+1$ ($m\leq m_{0}-1$), so it is indeed an
eigenstate on the Fermi level of uniform graphene, although it only exists in
the half plane $m\geq m_{0}$ ($m\leq m_{0}$). These \textquotedblleft
abnormal\textquotedblright\ evanescent eigenstates originate from the
singularity of the 2$\times$2 hopping matrix between neighboring unit cells.
For example, the wave function $|\Phi_{0}(m|m_{0})\rangle$ is nonzero on the
$A$ site of the unit cell $m=m_{0}$, but this site cannot hop to the
neighboring unit cell $m_{0}+1$ on its right (see Fig. \ref{G_PNJ}), so
$|\Phi_{0}(m|m_{0})\rangle$ vanishes for $m\geq m_{0}+1$. Similarly, the wave
function $|\Phi_{\infty}(m|m_{0})\rangle$ is nonzero on the $B$ site of the
unit cell $m=m_{0}$, but this site cannot hop to the neighboring unit cell
$m_{0}-1$ on its left, so $|\Phi_{\infty}(m|m_{0})\rangle$ vanishes for $m\leq
m_{0}-1$.

The \textquotedblleft abnormal\textquotedblright\ evanescent eigenstates do
not exist in an infinite uniform graphene, but they do appear near the
interface of the graphene junction. When considering the transmission of an
incident traveling wave, it is important to include these \textquotedblleft
abnormal\textquotedblright\ eigenstates.

\subsection{Mode-matching across P-N\ junctions}

Due to the translational invariance along the $Y$ axis, the scattering of
incident states with different $k_{Y}$'s is decoupled, so we need only
consider the 1D chain with a \textit{given} $k_{Y}\in\lbrack0,2\pi]$. As shown
in Fig. \ref{G_PNJ}, each unit cell contains two orbitals $A$ and $B$. The
2$\times2$ on-site energy of unit cell $m$ is $V_{m}\mathbf{I}_{2\times
2}+\mathbf{h}(k_{Y})$, where $V_{m}$ is the on-site potential: $V_{m}%
=V_{\mathrm{L}}$ for $m\leq0$ and $V_{m}=V_{\mathrm{R}}$ for $m\geq1$ [see
Fig. \ref{G_PNJ}(b)]. The 2$\times$2 hopping from unit cell $m+1$ to $m$ is
$\mathbf{t}$ [Eq. (\ref{T})]. The scattering state on the Fermi level $E_{F}$
satisfies the Schr\"{o}dinger equation for the 1D chain:
\begin{equation}
-\mathbf{t}^{\dag}|\Psi(m-1)\rangle+(E_{F}-\mathbf{h}(k_{Y})-V_{m}%
\mathbf{)}|\Psi(m)\rangle-\mathbf{t}|\Psi(m+1)\rangle=0. \label{SEQ_ALL}%
\end{equation}
Next we follow similar procedures to the usual mode-matching technique for
continuum models.

As the first step, we need to find all the left-going and right-going
eigenstates of each uniform region on the Fermi level. Here right-going
(left-going) eigenstates refer to both evanescent eigenstates that decay along
the $+X$ ($-X$) axis and traveling eigenstates whose group velocity $\partial
E(\mathbf{k})/\partial k_{X}$ along the $+X$ axis of the tilted coordinates,
or equivalently the group velocity $\partial E(\mathbf{k})/\partial k_{x}$
along the $+x$ axis of the Cartesian coordinates, is positive (negative)
\cite{AndoPRB1991}, where we have used $\partial/\partial k_{X}=(2/3)\partial
/\partial k_{x}$ according to Eq. (\ref{KXKY}). These eigenstates can be
obtained by exactly the same way as the previous subsection, with the only
difference being $E_{F}\rightarrow\varepsilon_{\mathrm{L}}=E_{F}%
-V_{\mathrm{L}}$ ($E_{F}\rightarrow\varepsilon_{\mathrm{R}}=E_{F}%
-V_{\mathrm{R}}$) for the left (right) region. For the region $s$
($=\mathrm{L}$ or $\mathrm{R}$), we obtain one right-going traveling
eigenstate and one left-going traveling eigenstate as the two intersection
points between the given horizontal line $k_{Y}$ and the Fermi contour
$|\varepsilon_{s}|=|f(\mathbf{k})|$ in the $\mathbf{k}=(k_{X},k_{Y}%
)_{\mathrm{T}}$ plane. For each region, we also obtain two \textquotedblleft
abnormal\textquotedblright\ evanescent eigenstates: Eqs. (\ref{EIG0})\ and
(\ref{EIGINF}). For clarity, we use $\lambda_{\alpha}\equiv e^{ik_{X,\alpha}}$
($\alpha=i,r,t$) for the propagation phases of different traveling eigenstates
over one unit cell along the $X$ axis: $\alpha=i$ ($\alpha=r)$ for the
right-going (left-going) eigenstate in the left N\ region and $\alpha=t$ for
the right-going traveling eigenstate in the right P region. The condition that
$\mathbf{k}_{i},\mathbf{k}_{r},$ and $\mathbf{k}_{t}$ lie on the Fermi contour
amounts to
\begin{subequations}
\label{ONSHELL}%
\begin{align}
\left\vert \varepsilon_{\mathrm{L}}\right\vert  &  =\left\vert f(\mathbf{k}%
_{i})\right\vert =\left\vert f(\mathbf{k}_{r})\right\vert ,\\
\left\vert \varepsilon_{\mathrm{R}}\right\vert  &  =\left\vert f(\mathbf{k}%
_{t})\right\vert,
\end{align}
where $f(\mathbf{k})$ is defined in Eq. (\ref{FK}). The wave vectors and spinors of the incident, reflection, and transmission
eigenstates are $\mathbf{k}_{\alpha}\equiv(k_{X,\alpha},k_{Y})_{\mathrm{T}}$
and $|\Phi_{\alpha}\rangle=[1,e^{i\varphi_{\alpha}}]^{T}$, where
\end{subequations}
\begin{subequations}
\label{PHI_ALPHA}%
\begin{align}
e^{i\varphi_{i}}  &  \equiv\frac{f^{\ast}(\mathbf{k}_{i})}{\left\vert
f(\mathbf{k}_{i})\right\vert },\\
e^{i\varphi_{r}}  &  \equiv\frac{f^{\ast}(\mathbf{k}_{r})}{\left\vert
f(\mathbf{k}_{r})\right\vert },\\
e^{i\varphi_{t}}  &  \equiv-\frac{f^{\ast}(\mathbf{k}_{t})}{\left\vert
f(\mathbf{k}_{t})\right\vert }.
\end{align}

As the second step, we consider a right-going incident traveling wave in the
left N\ region
\end{subequations}
\[
|\Phi_{i}(m)\rangle=e^{i(m-1)k_{X,i}}|\Phi_{i}\rangle
\]
and calculate the scattering state. By solving Eq. (\ref{SEQ_ALL}) for
$m=-\infty,\cdots,0$, we obtain the scattering state in the left N\ region as
the sum of the incident wave and the reflection wave:
\begin{equation}
|\Psi(m)\rangle=|\Phi_{i}(m)\rangle+|\Phi_{r}(m)\rangle\ \ (m\leq1),
\label{PHI_L}%
\end{equation}
where%
\begin{equation}
|\Phi_{r}(m)\rangle=re^{i(m-1)k_{X,r}}|\Phi_{r}\rangle+\tilde{r}|\Phi_{\infty
}(m|1)\rangle\label{PHIR}%
\end{equation}
is a linear combination of the left-going traveling eigenstate and the
left-going \textquotedblleft abnormal\textquotedblright\ evanescent eigenstate
in the left N\ region. By solving Eq. (\ref{SEQ_ALL}) for $m=1,2,\cdots
,+\infty$, we obtain the scattering state in the right P region as the
transmission wave%
\begin{equation}
|\Psi(m)\rangle=|\Phi_{t}(m)\rangle\ \ (m\geq0), \label{PHI_R}%
\end{equation}
which is a linear combination of the right-going traveling eigenstates and the
right-going \textquotedblleft abnormal\textquotedblright\ evanescent
eigenstate in the right P region:
\begin{equation}
|\Phi_{t}(m)\rangle=te^{imk_{X,t}}|\Phi_{t}\rangle+\tilde{t}|\Phi
_{0}(m|0)\rangle. \label{PHIT}%
\end{equation}
There are four unknown coefficients, including two reflection coefficients
$r,\tilde{r}$ and two transmission coefficients $t,\tilde{t}$.

As the final step, the overlap of Eqs. (\ref{PHI_L}) and (\ref{PHI_R}) at the
interface region $m=0,1$ gives the continuity condition $|\Phi_{i}%
(m)\rangle+|\Phi_{r}(m)\rangle=|\Phi_{t}(m)\rangle$ for $m=0,1$ [see the
dashed square in Fig. \ref{G_PNJ}(b)]. The equation at $m=1$ gives
\begin{subequations}
\label{EQM1}%
\begin{align}
1+r  &  =e^{ik_{X,t}}t,\label{EQM1A}\\
e^{i\varphi_{i}}+e^{i\varphi_{r}}r+\sqrt{2}\tilde{r}  &  =e^{ik_{X,t}%
}e^{i\varphi_{t}}t. \label{EQM1B}%
\end{align}
The equation at $m=0$ gives
\end{subequations}
\begin{subequations}
\label{EQM0}%
\begin{align}
e^{-ik_{X,i}}+e^{-ik_{X,r}}r  &  =t+\sqrt{2}\tilde{t},\label{EQM0A}\\
e^{-ik_{X,i}}e^{i\varphi_{i}}+e^{-ik_{X,r}}e^{i\varphi_{r}}r  &
=e^{i\varphi_{t}}t. \label{EQM0B}%
\end{align}
These four equations uniquely determine the four unknown coefficients
$r,\tilde{r},t,\tilde{t}$. Remarkably, Eqs. (\ref{EQM1A}) and (\ref{EQM0B})
form a closed set of equations for the \textit{traveling} eigenstates alone,
from which we obtain
\end{subequations}
\begin{subequations}
\label{RT}%
\begin{align}
r  &  =-\frac{e^{i(\varphi_{i}-k_{X,i})}-e^{i(\varphi_{t}-k_{X,t})}%
}{e^{i(\varphi_{r}-k_{X,r})}-e^{i(\varphi_{t}-k_{X,t})}},\\
t  &  =e^{-ik_{X,t}}(1+r).
\end{align}
Moreover, the \textquotedblleft abnormal\textquotedblright\ evanescent
eigenstates decay to zero over a single unit cell and hence do not contribute
to the reflection and transmission waves away from the interface; e.g.,
$|\Phi_{r}(m)\rangle=re^{i(m-1)k_{X,r}}|\Phi_{r}\rangle$ at $m\leq0$ only
contains the left-going traveling eigenstate, while $|\Phi_{t}(m)\rangle
=te^{imk_{X,t}}|\Phi_{t}\rangle$ at $m\geq1$ only contain the right-going
traveling eigenstates. Therefore, the presence of the \textquotedblleft
abnormal\textquotedblright\ evanescent eigenstates has \textit{no} effect on
the scattering properties of a sharp interface along the zigzag axis. The
transmission probability of an incident traveling eigenstate with wave vector
$k_{Y}$ on the Fermi level is
\end{subequations}
\begin{equation}
T(k_{Y})=1-|r(k_{Y})|^{2}. \label{TKY}%
\end{equation}
Perfect transmission corresponds to $T(k_{Y})=1$, which amounts to
$r(k_{Y})=0$ or equivalently%
\begin{equation}
e^{i(\varphi_{i}-k_{X,i})}=e^{i(\varphi_{t}-k_{X,t})}\neq e^{i(\varphi
_{r}-k_{X,r})}. \label{MC}%
\end{equation}

\begin{figure}[pth]
\includegraphics[width=\columnwidth,clip]{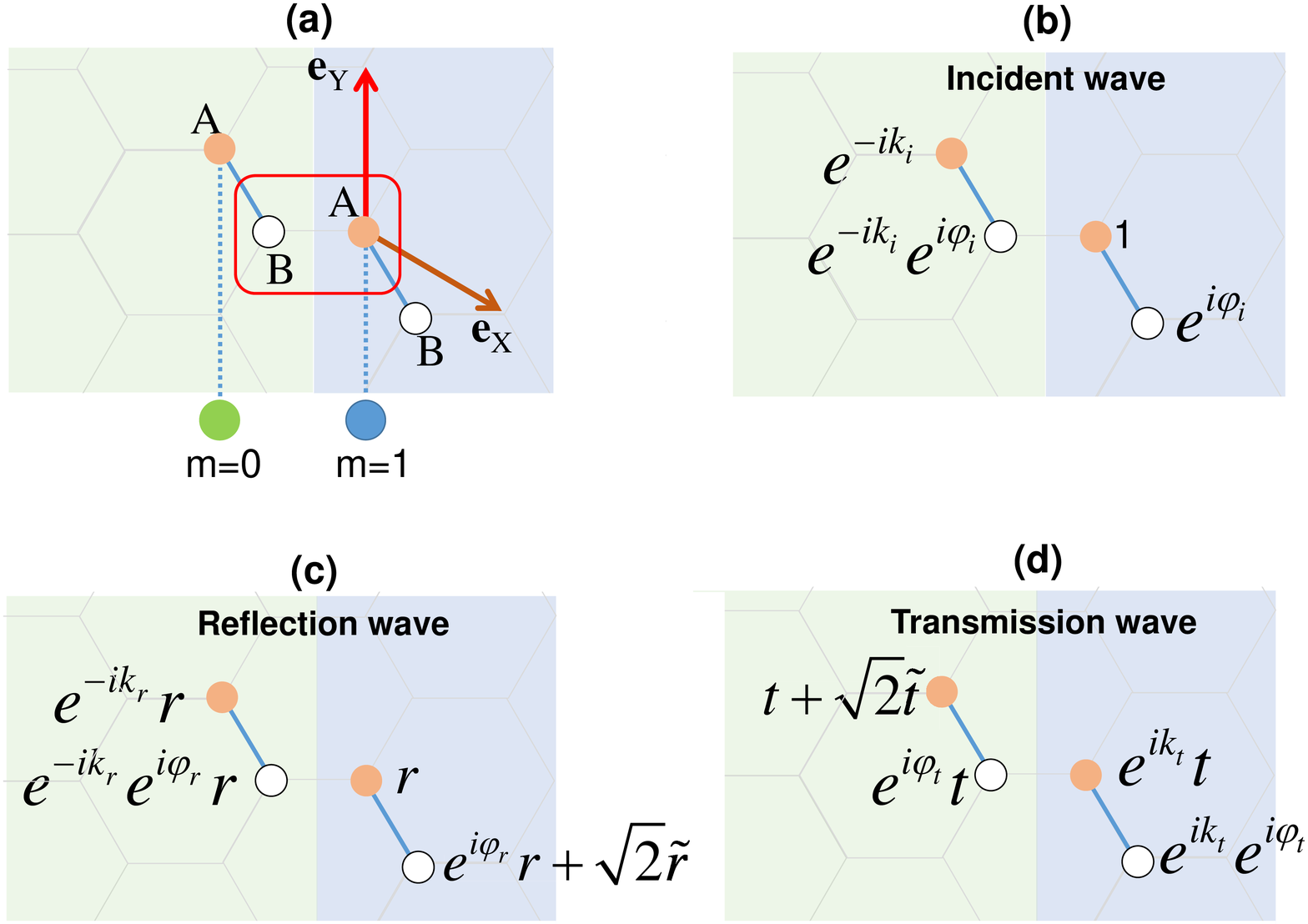}\caption{(a) Continuity
of the scattering wave function at the interface unit cells $m=0$ and $1$.
Each unit cell contains two atoms, as connected by the thick solid line.
(b)-(d) show the amplitudes of the incident, reflection, and transmission
waves at each carbon site.}%
\label{G_MATCHING}%
\end{figure}

Next we discuss the mode-matching condition at the interface $m=0,1$ (see Fig.
\ref{G_MATCHING}) in more detail to explain why the \textquotedblleft
abnormal\textquotedblright\ evanescent eigenstates do not affect the
transmission of the traveling eigenstate. The key is that the 2$\times$1
spinor $|\Phi\rangle=[a,b]^{T}$ means that the wave amplitude at the $A$ site
is $a$, while that at the $B$ site is $b$. Therefore, the matching of the
2$\times$1 spinor wave function at $m=0$ and $1$ amounts to the matching of
the wave amplitudes at the four sites in Fig. \ref{G_MATCHING}:\ Eq.
(\ref{EQM1A}) [Eq. (\ref{EQM1B})] for the matching at the $A$ ($B$) site of
unit cell $m=1$ and Eq. (\ref{EQM0A}) [Eq. (\ref{EQM0B})] for the matching at
the $A$ ($B$) site of $m=0$. However, the left-going ideal evanescent state
$|\Phi_{\infty}(m|1)\rangle$ contained in $|\Phi_{r}(m)\rangle$ is nonzero
only on the $B$ site of $m=1$, while the right-going ideal evanescent
eigenstate $|\Phi_{0}(m|0)\rangle$ contained in $|\Phi_{t}(m)\rangle$ is
nonzero only on the $A$ site of $m=0$. Therefore, the matching conditions on
the $A$ site of $m=1$ [Eq. (\ref{EQM1A})] and on the $B$ site of $m=0$ [Eq.
(\ref{EQM0B})], as enclosed by the red box in Fig. \ref{G_MATCHING}, do not
contain any \textquotedblleft abnormal\textquotedblright\ evanescent waves.
The matching of the traveling waves at these two sites uniquely determines the
reflection and transmission coefficients $r$ and $t$ for the traveling eigenstates.

\subsection{Generalized pseudospin in tight-binding model}

\label{SEC_PSEUDOSPIN}

In the linear continuum model, the perfect Klein tunneling at normal incidence
has a physically transparent interpretation as the conservation of the
pseudospin \cite{KatsnelsonNatPhys2006,BeenakkerRMP2008,CulcerPRB2008}. In the
tight-binding model, however, such a simple physical picture for the Klein
tunneling is still lacking, due to the broken chirality of the Dirac fermions
by the trigonal warping
\cite{AjikiJPSJ1996,AndoJPSJ1998,WuPRL2007a,DombrowskiPRL2017}.

Here we demonstrate that in the tight-binding model, the perfect transmission
over the energy range $E\in[-1,+1]$ can always be interpreted as the
conservation of a \textit{generalized} pseudospin. The key is that the
mode-matching conditions for the traveling waves at the interface form a
closed set of equations [Eqs. (\ref{EQM1A}) and (\ref{EQM0B})], which can be
put into the form
\begin{equation}
|u_{i}\rangle+r|u_{r}\rangle=te^{ik_{X,t}}|u_{t}\rangle, \label{CC_TB}%
\end{equation}
where
\[
|u_{\alpha}\rangle\equiv%
\begin{bmatrix}
1\\
e^{i(\varphi_{\alpha}-k_{X,\alpha})}%
\end{bmatrix}
\ \ (\alpha=i,r,t).
\]
The perfect transmission condition [Eq. (\ref{MC})] leads to $|u_{i}%
\rangle=|u_{t}\rangle\neq|u_{r}\rangle$. Conversely, once $|u_{i}%
\rangle=|u_{t}\rangle\neq|u_{r}\rangle$, we immediately obtain $r=0$ and hence
perfect transmission. Thus we arrive at the following \textit{necessary and
sufficient condition} for perfect transmission:%
\begin{equation}
|u_{i}\rangle=|u_{t}\rangle\neq|u_{r}\rangle. \label{PT_TB}%
\end{equation}
By regarding the $A,B$ sites inside the red box of Fig. \ref{G_MATCHING} as a
special unit cell across the interface, $|u_{i}\rangle$, $r|u_{r}\rangle$, and
$te^{ik_{X,t}}|u_{t}\rangle$ become the amplitudes of the incident,
reflection, and transmission waves inside this unit cell, as shown in Fig.
\ref{G_MATCHING}(b)-(d). Therefore, Eq. (\ref{CC_TB}) simply expresses the
continuity of the scattering wave function inside this special unit cell. Then
we can associate each spinor with a generalized pseudospin
\begin{equation}
\boldsymbol{\sigma}_{\alpha}\equiv\frac{\langle u_{\alpha}|\boldsymbol{\hat
{\sigma}}|u_{\alpha}\rangle}{\langle u_{\alpha}|u_{\alpha}\rangle}%
=(\cos(\varphi_{\alpha}-k_{X,\alpha})\mathbf{,}\sin(\varphi_{\alpha
}-k_{X,\alpha})), \label{TB_SIMGA}%
\end{equation}
where $\boldsymbol{\hat{\sigma}}=(\hat{\sigma}_x,\hat{\sigma}_y)$ and $\hat{\sigma}_{x,y}$ are Pauli matrices. Then the perfect
transmission condition amounts to
\begin{equation}
\boldsymbol{\sigma}_{i}=\boldsymbol{\sigma}_{t}\neq\boldsymbol{\sigma}_{r},
\label{PT_SIGMA}%
\end{equation}
i.e., conservation of the generalized pseudospin across the graphene junctions.

\subsection{Generalizations to finite-width junctions}

\begin{figure}[pth]
\includegraphics[width=\columnwidth,clip]{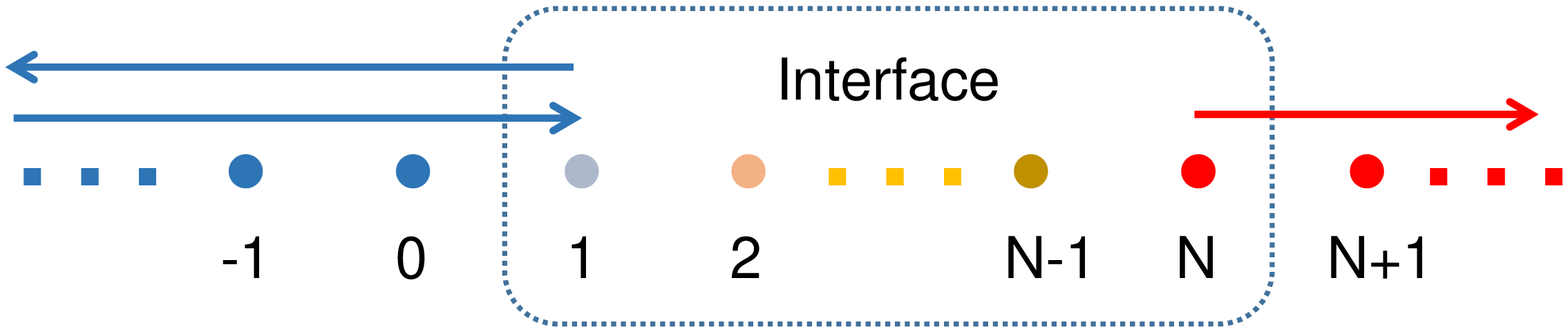}\caption{An
arbitrary junction consists of the left uniform region ($m\leq0)$, the
interface region ($1\leq m\leq N$), and the right uniform region ($m\geq
N+1)$. The arrows indicate the incident wave, the reflection wave, and the
transmission wave.}%
\label{G_GENERALIZATION}%
\end{figure}

The mode-matching technique in the tilted coordinates can also deal with
finite-width junctions. Suppose the left N\ region, the right P region, and
the interface region consist of the unit cells $(m,n)$ with $m\leq0$, $m\geq
N+1$, and $1\leq m\leq N$, respectively. By using the Fourier transform Eq.
(\ref{FT_X}) from the on-site basis into the hybrid basis $|m,k_{Y}%
,\alpha\rangle$, we reduce the 2D junction into a 1D chain parameterized by a
given $k_{Y}$. Each unit cell contains two orbitals $A$ and $B$. The
2$\times2$ on-site energy of the unit cell $m$ is $V_{m}\mathbf{I}_{2\times
2}+\mathbf{h}(k_{Y})$, where the on-site potential $V_{m}=V_{\mathrm{L}}$ in
the left N region ($m\leq0$), $V_{m}=V_{\mathrm{R}}$ in the right P region
($m\geq N+1$), and $V_{m}$ can be arbitrary in the interface region ($1\leq
m\leq N$), as shown in Fig. \ref{G_GENERALIZATION}(b). The 2$\times$2 hopping
from unit cell $m+1$ to $m$ is $\mathbf{t}$ [Eq. (\ref{T})].

For a right-going incident wave on the Fermi level from the left N region
$|\Phi_{i}(m)\rangle=e^{i(m-1)k_{X,i}}|\Phi_{i}\rangle$, the scattering state
is obtained by solving the 1D Schr\"{o}dinger equation Eq. (\ref{SEQ_ALL}).
Solving Eq. (\ref{SEQ_ALL}) with $m\leq0$ gives
\[
|\Psi(m)\rangle=|\Phi_{i}(m)\rangle+re^{i(m-1)k_{X,r}}|\Phi_{r}\rangle
+\tilde{r}\lambda_{\infty}^{(m-1)}|\Phi_{\infty}\rangle\ \ (m\leq1),
\]
where the last two terms are reflection waves. Solving Eq. (\ref{SEQ_ALL})
with $m\geq N+1$ gives the transmission waves
\[
|\Psi(m)\rangle=te^{i(m-N)k_{X,t}}|\Phi_{X,t}\rangle+\tilde{t}\lambda
_{0}^{(m-N)}|\Phi_{0}\rangle\ (m\geq N).
\]
The unknown variables$r,\tilde{r},t,\tilde{t}$, and $|\Psi(m)\rangle$ ($2\leq
m\leq N-1$) can be uniquely determined by Eq. (\ref{SEQ_ALL}) with
$m=1,2,\cdots,N$. This
is reminiscent of the mode-matching method developed by Ando
\cite{AndoPRB1991}, albeit in the tilted coordinates. For large $N$, analytical solutions are no longer available, and numerical calculations are necessary.

\section{Numerical results and discussions}

\begin{figure}[pth]
\includegraphics[width=\columnwidth,clip]{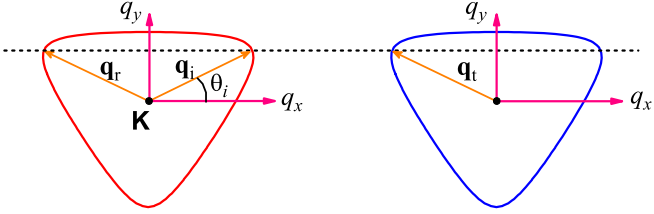}\caption{Fermi contours
for the N region (conduction band, red contour) and P\ region (valence band,
blue contour) of a symmetric P-N junction with $\varepsilon_{\mathrm{L}%
}=\varepsilon_{\mathrm{R}}=0.9$. The arrows indicate the reduced wave vectors
$\mathbf{q}_{i},\mathbf{q}_{r}$, and $\mathbf{q}_{t}$ for the incident,
reflection, and transmission waves.}%
\label{G_CONTOUR}%
\end{figure}

The mirror symmetry of the P-N junction about the $x$ axis guarantees the
transmission probability $T(k_{y})$ to be an even function of $k_{y}$. So we
limit our discussions to one valley, say $\mathbf{K}$, and define the reduced
wave vector $\mathbf{q}\equiv\mathbf{k}-\mathbf{K}$. For specificity we
calculate the transmission probability of electrons in the $\mathbf{K}$ valley
across a symmetric graphene P-N junction with $\varepsilon_{\mathrm{L}%
}=-\varepsilon_{\mathrm{R}}=V_{0}$; i.e., the electron doping in the left N
region is equal to the hole doping in the right P region. We present our
calculation results in the conventional Cartesian coordinates (see Fig.
\ref{G_GRAPHENE}) to make them accessible to readers that are more familiar
with the Cartesian coordinates. For a given $q_{y}$, the wave vectors of the
incident, reflection, and transmission waves correspond to the intersection
points of the horizontal line $q_{y}$ and the Fermi contour in the
$(q_{x},q_{y})$ plane, as shown in Fig. \ref{G_CONTOUR}. The incident wave is
characterized by either $q_{y}$ or the azimuth angle $\theta_{i}$ of
$\mathbf{q}_{i}$, so the transmission probability is denoted by $T(q_{y})$ or
$T(\theta_{i})$. For clarity we define $q_{y,\mathrm{PT}}$ ($\theta
_{\mathrm{PT}}$)\ as the critical momentum (angle) leading to perfect
transmission: $T(q_{y,\mathrm{PT}})=1$ [$T(\theta_{\mathrm{PT}})=1$]. For
reference, the conventional continuum model with a linear dispersion gives the
transmission probability $T(\theta_{i})=\cos^{2}\theta_{i}$ (Ref.
\onlinecite{CheianovPRB2006}) and hence $q_{y,\mathrm{PT}}=\theta
_{\mathrm{PT}}=0$.

\subsection{Numerical results}

\begin{figure}[pth]
\includegraphics[width=\columnwidth,clip]{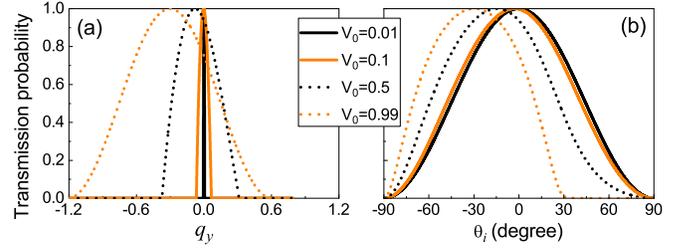}\caption{(a)
Transmission probability as a function of $q_{y}$. (b) Transmission
probability as a function of the azimuth angle $\theta_{i}$ of the incident
wave vector $\mathbf{q}_{i}$.}%
\label{G_OBLIQUE}%
\end{figure}

At the beginning, we plot in Fig. \ref{G_OBLIQUE} the exact results from the
tight-binding model for the transmission probability, as calculated from Eqs.
(\ref{TKY}) and (\ref{RT}). At very low doping $V_{0}=0.01$, perfect
transmission occurs at $q_{y,\mathrm{PT}}=\theta_{\mathrm{PT}}=0$, in
agreement with the well-known Klein tunneling in the widely used linear
continuum model \cite{KatsnelsonNatPhys2006}. However, with increasing doping
level, $q_{y,\mathrm{PT}}$ and $\theta_{\mathrm{PT}}$ exhibit larger and
larger deviations from zero, indicating perfect transmission at oblique incidence.

\begin{figure}[pth]
\includegraphics[width=\columnwidth,clip]{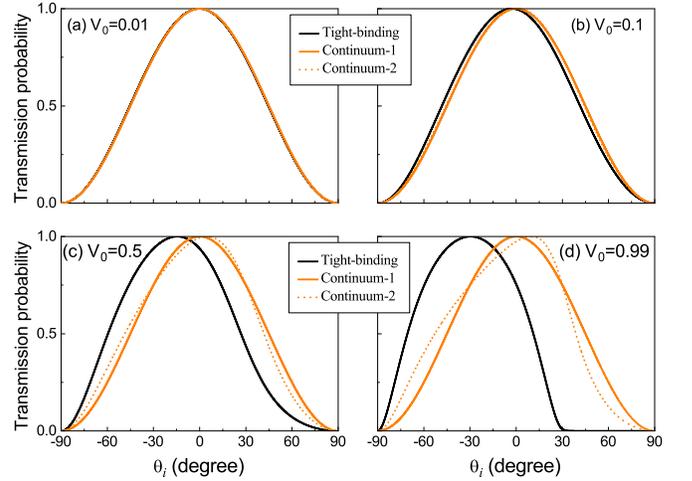}\caption{Transmission
probability from the tight-binding model vs those from the linear continuum
model (solid orange line) or with quadratic corrections (dashed orange line)
for different doping levels: (a) $V_{0}=0.01$, (b) $V_{0}=0.1$, (c)
$V_{0}=0.5$, and (d) $V_{0}=0.99$.}%
\label{G_COMPARE}%
\end{figure}

In Fig. \ref{G_COMPARE}, we compare the exact results from the tight-binding
model to those from the continuum models. At very low doping $V_{0}=0.01$
[Fig. \ref{G_COMPARE}(a)], all the models agree with each other and give
$\theta_{\mathrm{PT}}=0$. However, with increasing doping level, the results
from the continuum models deviate more and more from the exact results.
Actually, the deviation is appreciable even at relatively low doping
$V_{0}=0.1$ [Fig. \ref{G_COMPARE}(b)], where the trigonal warping is small:
$\eta=3\%$ [Fig. \ref{G_GRAPHENE}(c)]. Interestingly, even the continuum model
with a quadratic correction (dashed orange line) to incorporate the trigonal
warping up to the lowest order (see the Appendix) does not agree with the exact results.

\begin{figure}[t]
\includegraphics[width=\columnwidth,clip]{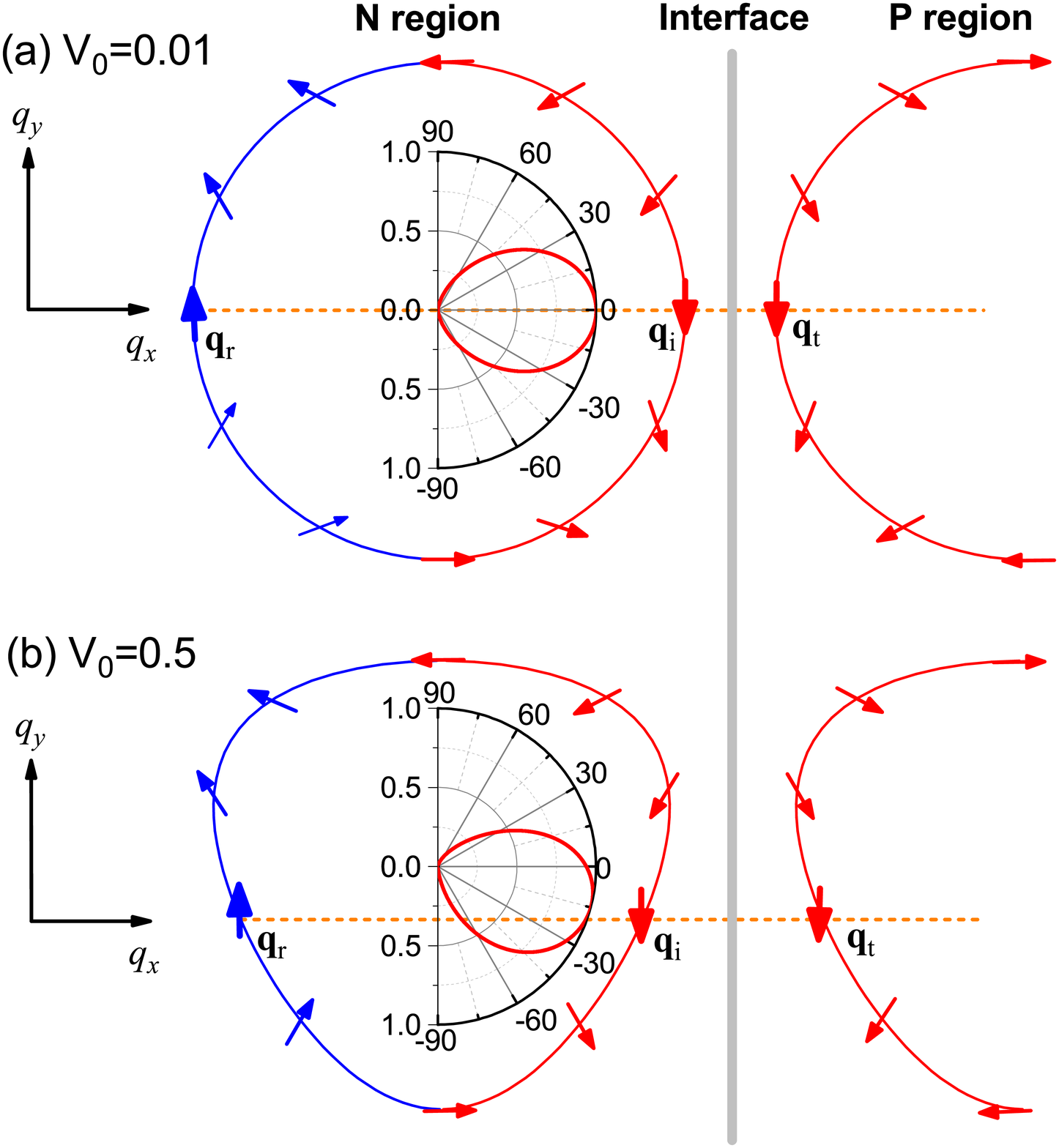}\caption{Fermi contours
for each region of the graphene P-N junction in the Cartesian coordinate
$(q_{x},q_{y})$ for different doping levels: (a) $V_{0}=0.01$ and (b)
$V_{0}=0.5$. Red (blue) for right-going (left-going) states and arrows for
their pseudospins. In each panel, the inset shows the transmission probability
$T(\theta_{i})$ as a function of the azimuth angle $\theta_{i}$ of the
incident wave vector $\mathbf{q}_{i}$.}%
\label{G_TEXTURE}%
\end{figure}

Next we illustrate the generalized pseudospin picture (see Sec.
\ref{SEC_PSEUDOSPIN}) in the tight-binding model. As shown in Fig.
\ref{G_TEXTURE}, for a given $q_{y}$, the wave vectors $\mathbf{q}%
_{i},\mathbf{q}_{r},\mathbf{q}_{t}$ of the incident, reflection, and
transmission waves are obtained as the intersection points between a
horizontal line (see the orange dashed line for an example) corresponding to
this $q_{y}$ and the Fermi contour. For low doping $V_{0}=0.01$, the condition
$\boldsymbol{\sigma}_{i}=\boldsymbol{\sigma}_{t}\neq\boldsymbol{\sigma}_{r}$
is satisfied at $q_{y}=0$, so perfect transmission occurs at normal incidence
$\theta_{\mathrm{PT}}=0$ [Fig. \ref{G_TEXTURE}(a)]. For higher doping
$V_{0}=0.5$ [Fig. \ref{G_TEXTURE}(b)], the texture of the generalized
pseudospins on the Fermi contour is twisted relative to those at low doping.
In this case, the condition $\boldsymbol{\sigma}_{i}=\boldsymbol{\sigma}%
_{t}\neq\boldsymbol{\sigma}_{r}$ occurs at $\theta_{\mathrm{PT}}\approx-14.6%
\operatorname{{{}^\circ}}%
$; i.e., perfect transmission occurs for oblique incidence.

\subsection{Experimental feasibility}

\begin{figure}[pth]
\includegraphics[width=\columnwidth,clip]{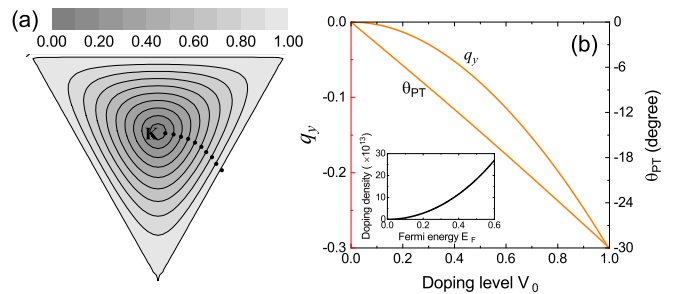}\caption{(a) Incident
wave vector (black dots) on the Fermi contour leading to perfect transmission
under different doping levels $V_{0}$. (b) Critical momentum $q_{y,\mathrm{PT}%
}$ and angle $\theta_{\mathrm{PT}}$ for perfect transmission as functions of
the doping level. The inset shows the doping density vs. the
Fermi energy, as calculated from the tight-binding model.}%
\label{critical}%
\end{figure}

Ever since the initial theoretical prediction \cite{KatsnelsonNatPhys2006}, many experimental efforts
\cite{HuardPRL2007,GorbachevNL2008,StanderPRL2009,YoungNatPhys2009,RossiPRB2010,SajjadPRB2012,SutarNL2012,RahmanAPL2015,GutierrezNatPhys2016,ChenScience2016,BaiPRB2017}
have been devoted to Klein tunneling in graphene, culminating in the
experimental demonstration of the prominent angular dependence of the
transmission probability in graphene P-N\ junctions
\cite{SajjadPRB2012,SutarNL2012,RahmanAPL2015,ChenScience2016}. In particular,
the recent fabrication of high-quality graphene P-N junctions with high doping
levels \cite{DombrowskiPRL2017} makes the high-energy transmission across
graphene P-N junctions experimentally accessible; e.g., the perfect
transmission angle $\theta_{\mathrm{PT}}$ may be extracted by measuring the
angular dependence of transmission probability. To serve future experiments,
we plot the perfect transmission momentum $q_{y,\mathrm{PT}}$ and the perfect
transmission angle $\theta_{\mathrm{PT}}$ as functions of the doping level
$V_{0}$ in Fig. \ref{critical}.

Up to now, we have focused on symmetric P-N\ junctions with a sharp interface
and equal doping level and trigonal warping on both sides. For asymmetric P-N
junctions, the trigonal warping in the N region is not equal to that in the P
region. This may weaken the effect of perfect transmission at oblique
incidence by shifting $\theta_{\mathrm{PT}}$ towards zero. A smooth junction
potential usually suppresses the transmission as the incident angle increases
\cite{CheianovPRB2006} and plays an important role in the electron optics in
the graphene P-N junction \cite{ChenScience2016}. This may hinder the
experimental observation of perfect transmission at oblique incidence.
Fortunately, a very recent experiment \cite{BaiArxiv2017} shows that
atomically sharp graphene P-N junctions can be fabricated on the copper
surface. The potential difference between the P and the N regions can reach
$2V_{0}=660$ meV, corresponding to a doping level $V_{0}=0.33$ eV and doping
density $10^{13}$ cm$^{-2}$ [see the inset of Fig. \ref{critical}(b)]. At this
doping level, the trigonal warping is $\eta=4\%$ and the perfect transmission
angle is $\theta_{\mathrm{PT}}\approx-3.5%
\operatorname{{{}^\circ}}%
$. Moreover, a previous theoretical study shows that the electron-electron
interaction \cite{RoldanPRB2008} generally enhances the trigonal warping.
Therefore, we expect the perfect transmission at oblique incidence to be
experimentally accessible in the near future.

Finally, we note the intensive experimental activities in simulating the
honeycomb lattices by cold atoms and optical lattices \cite{PoliniNatNano2013}%
. In particular, the possibility of Klein tunneling in these system has been
examined
\cite{LonghiPRB2010,CasanovaPRA2010,ZhangPRA2012,FangPRB2016,OzawaPRA2017} and
observed recently \cite{GerritsmaPRL2011,SalgerPRL2011}. These simulated
graphene systems may provide an alternative platform for observing the perfect
transmission at oblique incidence. It would also be interesting to explore the
effect of trigonal warping on the transmission across bilayer graphene P-N
junctions \cite{TudorovskiyPS2012,CulcerPRB2009a}.

\section{Conclusions}

We have developed an analytical mode-matching technique to study the electron
transmission across graphene P-N junctions over a wide energy range. In
contrast to the well-known Klein tunneling at normal incidence for low
energies in the linear Dirac regime, we find that at energies beyond the
linear Dirac regime, the Klein tunneling at normal incidence becomes imperfect
and trigonal warping causes perfect transmission at oblique incidence. We show
that this phenomenon arises from the conservation of a generalized pseudospin
in the tight-binding model. This generalizes the well-known pseudospin picture
for perfect Klein tunneling, previously limited to low energies in the linear
Dirac regime, to all the energy ranges. The perfect transmission at oblique
incidence cannot be described by the continuum model, even after quadratic
corrections have been introduced to incorporate trigonal warping up to the
leading order. Our work may be relevant for the applications of the graphene P-N
junction in electronics and electron optics.

\section*{Acknowledgements}

This work was supported by the National Key R$\&$D Program of China (Grant
No. 2017YFA0303400), the NSFC (Grants No. 11504018, No. 11774021, No.
11274036, and No. 11322542), the MOST of China (Grant No. 2014CB848700), and the NSFC
program for ``Scientific Research Center''\ (Grant No. U1530401). S.H.Z.
thanks the preliminary research program (YY1708) of the College of Science of
BUCT. We acknowledge the computational support from the Beijing Computational
Science Research Center (CSRC).

\appendix

\section{Hamiltonian of uniform graphene in Cartesian coordinates}

\begin{figure}[pth]
\includegraphics[width=\columnwidth,clip]{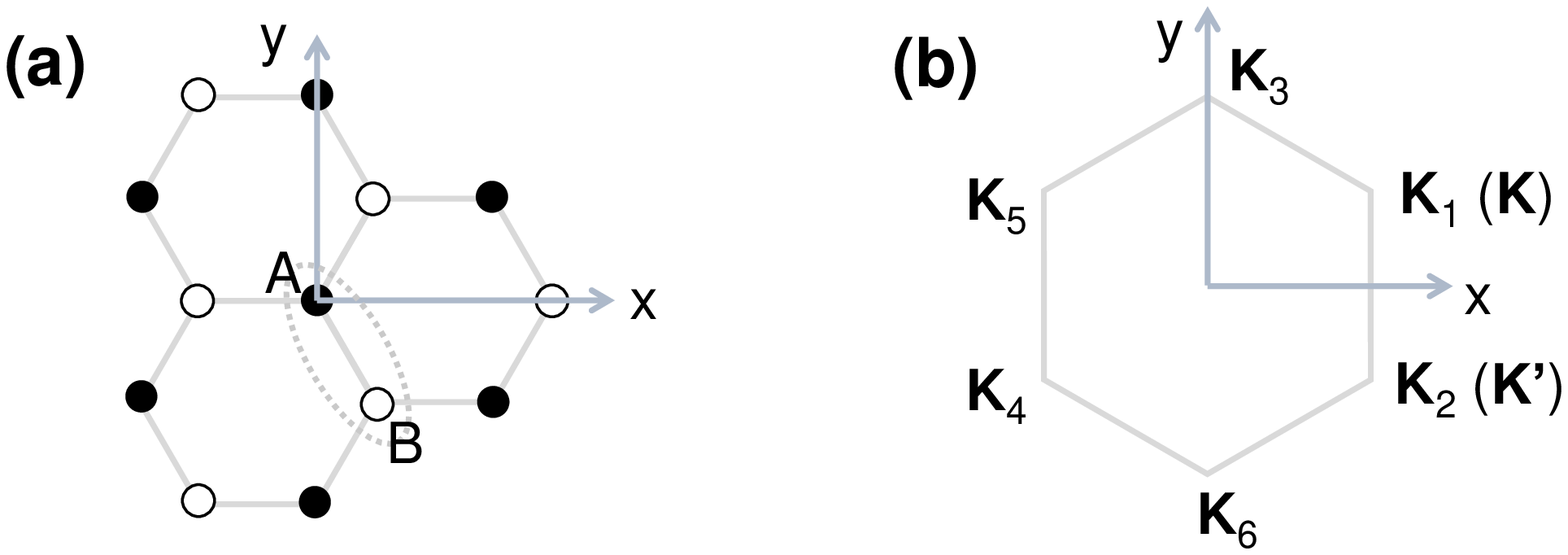}\caption{(a) Lattice
structure of graphene:\ filled (empty) circles for sublattice $A$ ($B$). (b)
Six Dirac points at the edge of the first Brillouin zone of graphene.}%
\label{G_APPEND1}%
\end{figure}

As shown in Fig. \ref{G_APPEND1}(a), the honeycomb lattice of graphene
consists of two sublattices (denoted by $A$ and $B$) and each unit cell
contains two carbon atoms (or $\pi_{z}$-orbitals), one on each sublattice. The
tight-binding Hamiltonian of graphene is \ \ \ \
\[
\hat{H}_{0}=-\sum_{\left\langle i,j\right\rangle }|i,A\rangle\langle
j,B|-h.c.,
\]
where $\langle i,j\rangle$ sums over all the nearest-neighbor carbon pairs,
$|i,\lambda\rangle$ ($\lambda=A,B$) is the $\pi_{z}$ orbital on the sublattice
$\lambda$ of unit cell $i$, and we have taken the nearest-neighbor hopping
energy as the unit of energy. We take the $A$ site as the origin of each unit
cell, so the relative displacements of the two carbon atoms inside the unit
cell are $\boldsymbol{\tau}_{A}=0$ and $\boldsymbol{\tau}_{B}=(1/2,-\sqrt
{3}/2)$, where we have taken the carbon-carbon bond length as the unit of
length. In terms of the unit cell location $\mathbf{R}_{i}$, the location of
the site $\lambda$ in unit cell $i$ is $\mathbf{R}_{i,\lambda}=\mathbf{R}%
_{i}+\boldsymbol{\tau}_{\lambda}$.

By Fourier transforming the real-space basis $\{|i,\lambda\rangle\}$ into the
momentum space basis $\{|\mathbf{k},\lambda\rangle\}$, the graphene
Hamiltonian can be transformed into the momentum space:%
\[
\hat{H}_{0}=\sum_{\mathbf{k}}f(\mathbf{k})|\mathbf{k},A\rangle\langle
\mathbf{k},B|+h.c.,
\]
where $f(\mathbf{k)}=-\sum_{i=1,2,3}e^{i\mathbf{k}\cdot\mathbf{d}_{i}}$ and
$\{\mathbf{d}_{i}\}$ depend on the choices on the Fourier transform. Namely,
if we make the Fourier transform using the location $\{\mathbf{R}_{i\lambda
}\}$ of the carbon \textit{sites}:
\[
|\mathbf{k},\lambda\rangle_{\mathrm{I}}\equiv\frac{1}{\sqrt{N}}\sum
_{i}e^{i\mathbf{k}\cdot\mathbf{R}_{i\lambda}}|i,\lambda\rangle,
\]
then $\mathbf{d}_{1},\mathbf{d}_{2},\mathbf{d}_{3}$ denote the relative
displacements of the three nearest-neighbor $B$ \textit{sites} [empty circles
in Fig. \ref{G_APPEND1}(a)] with respect to the central $A$ \textit{site}
[blue filled circle in Fig. \ref{G_APPEND1}(a)]:%
\[
\mathbf{d}_{1}^{\mathrm{I}}=\boldsymbol{\tau}_{B},\ \ \mathbf{d}%
_{2}^{\mathrm{I}}=\left(  \frac{1}{2},\frac{\sqrt{3}}{2}\right)
,\ \ \mathbf{d}_{3}^{\mathrm{I}}=\left(  -1,0\right)  .
\]
However, if we make the Fourier transform using the location $\{\mathbf{R}%
_{i}\}$ of the \textit{unit cells}:%

\[
|\mathbf{k},\lambda\rangle_{\mathrm{II}}\equiv\frac{1}{\sqrt{N}}\sum
_{i}e^{i\mathbf{k}\cdot\mathbf{R}_{i}}|i,\lambda\rangle,
\]
then $\mathbf{d}_{1},\mathbf{d}_{2},\mathbf{d}_{3}$ denote the relative
displacements of the three nearest-neighbor \textit{unit cells} [black filled
circles in Fig. \ref{G_APPEND1}(a)] with respect to the central \textit{unit
cell }[blue filled circle in Fig. \ref{G_APPEND1}(a)]:
\[
\mathbf{d}_{1}^{\mathrm{II}}=0,\ \ \ \mathbf{d}_{2}^{\mathrm{II}}=(0,\sqrt
{3})\text{,}\ \ \ \mathbf{d}_{3}^{\mathrm{II}}=\left(  -\frac{3}{2}%
,\frac{\sqrt{3}}{2}\right)  .
\]
These two choices are connected by $|\mathbf{k},A\rangle_{\mathrm{II}%
}=|\mathbf{k},A\rangle_{\mathrm{I}}$, $|\mathbf{k},B\rangle_{\mathrm{II}%
}=e^{-i\mathbf{k}\cdot\boldsymbol{\tau}_{B}}|\mathbf{k},B\rangle_{\mathrm{I}}%
$, $\mathbf{d}_{i}^{\mathrm{II}}=\mathbf{d}_{i}^{\mathrm{I}}-\boldsymbol{\tau
}_{B}$, and $f_{\mathrm{II}}(\mathbf{k})=$ $e^{-i\mathbf{k}\cdot
\boldsymbol{\tau}_{B}}f_{\mathrm{I}}(\mathbf{k})$. Both cases satisfy
$f^{\ast}(-\mathbf{k})=f(\mathbf{k})$; thus under time-reversal operation
$\hat{\theta}$, which leaves $|i\lambda\rangle$ invariant and brings
$|\mathbf{k},\lambda\rangle$ into $\hat{\theta}|\mathbf{k},\lambda
\rangle=|-\mathbf{k,}\lambda\rangle$, the graphene Hamiltonian remains
invariant:
\[
\hat{\theta}\hat{H}_{0}\hat{\theta}^{-1}=\sum_{\mathbf{k}}f^{\ast}%
(\mathbf{k})|-\mathbf{k},A\rangle\langle-\mathbf{k},B|+h.c.=\hat{H}_{0}.
\]
Diagonalizing the Hamiltonian in the momentum space gives one conduction band
and one valence band $E_{\pm}(\mathbf{k})=\pm|f(\mathbf{k})|$ or explicitly
\begin{equation}
E_{\pm}(\mathbf{k})=\pm\sqrt{3+2\cos(\sqrt{3}k_{y})+4\cos\frac{3k_{x}}{2}%
\cos\frac{\sqrt{3}k_{y}}{2}}, \label{CMED}%
\end{equation}
which touch each other (i.e., $f(\mathbf{k})=0$) at six Dirac points at the
edge of the first Brillouin zone [Fig. \ref{G_APPEND1}(b)]. Since
$\mathbf{K}_{1},\mathbf{K}_{3},\mathbf{K}_{5}$ ($\mathbf{K}_{2},\mathbf{K}%
_{4},\mathbf{K}_{6}$) only differ by a reciprocal vector, only two Dirac
points are inequivalent, e.g., $\mathbf{K}_{1}$ and $\mathbf{K}_{2}$.
Hereafter we denote $\mathbf{K}_{1}$ as $\mathbf{K}$ and $\mathbf{K}_{2}$ as
$\mathbf{K}^{\prime}$:
\[
\mathbf{K}=\frac{2\pi}{3}\left(  1,\frac{1}{\sqrt{3}}\right)  ,\mathbf{K}%
^{\prime}=\frac{2\pi}{3}\left(  1,-\frac{1}{\sqrt{3}}\right)  .
\]

The continuum model near the Dirac point $\mathbf{K}_{j}$ is obtained by
considering $\mathbf{k}\approx\mathbf{K}_{j}$ and expanding $f(\mathbf{k})$
into Taylor series of $\mathbf{q}\equiv\mathbf{k}-\mathbf{K}_{j}$:%
\[
\hat{H}_{j}(\mathbf{q})=\sum_{\mathbf{q}}f_{j}(\mathbf{q})|\mathbf{K}%
_{j},A\rangle\langle\mathbf{K}_{j},B|+h.c.,
\]
where $f_{j}(\mathbf{q})\equiv f(\mathbf{K}_{j}+\mathbf{q})$. Up to the second
order of $\mathbf{q}$, we have%
\begin{align*}
f_{1,\mathrm{I}}(\mathbf{q})  &  =f_{4,\mathrm{I}}^{\ast}(-\mathbf{q}%
)=\frac{3}{2}qe^{-i\pi/6+i\theta_{\mathbf{q}}}-\frac{3}{8}q^{2}e^{i\pi
/3}e^{-i2\theta_{\mathbf{q}}},\\
f_{2,\mathrm{I}}(\mathbf{q})  &  =f_{5,\mathrm{I}}^{\ast}(-\mathbf{q}%
)=\frac{3}{2}qe^{-i\pi/6\mathbf{-}i\theta_{\mathbf{q}}}-\frac{3}{8}%
q^{2}e^{i\pi/3}e^{i2\theta_{\mathbf{q}}},\\
f_{3,\mathrm{I}}(\mathbf{q})  &  =f_{6,\mathrm{I}}^{\ast}(-\mathbf{q}%
)=\frac{3}{2}iqe^{i\theta_{\mathbf{q}}}+\frac{3}{8}q^{2}e^{-i2\theta
_{\mathbf{q}}},
\end{align*}
for choice I, where $\theta_{\mathbf{q}}$ is the azimuthal angle of
$\mathbf{q}$. This choice gives a concise form for the $O(q^{2})$ terms, so it
is usually used to study the trigonal warping effect of graphene
\cite{ReijndersPRB2017}. For choice II, the $O(q^{2})$ term is very
complicated (although different choices of the Fourier transform give the same
physics \cite{BenaNJP2009}), so we only give the Taylor expansion up to
$O(q)$:
\begin{align*}
f_{1,\mathrm{II}}(\mathbf{q})  &  =f_{3,\mathrm{II}}(\mathbf{q}%
)=f_{5,\mathrm{II}}(\mathbf{q})=\frac{3}{2}qe^{-i\pi/6+i\theta_{\mathbf{q}}%
},\\
f_{2,\mathrm{II}}(\mathbf{q})  &  =f_{4,\mathrm{II}}(\mathbf{q}%
)=f_{6,\mathrm{II}}(\mathbf{q})=-\frac{3}{2}qe^{i\pi/6-i\theta_{\mathbf{q}}}.
\end{align*}
For either case, we have the identity $f_{i+3}(\mathbf{q})=f_{i}^{\ast
}(-\mathbf{q})$, which follows from the time-reversal symmetry $f^{\ast
}(-\mathbf{k})=f(\mathbf{k})$.


\end{document}